\documentclass[twocolumn,aps,epsfig,nofootinbib,preprintnumbers]{revtex4}

\usepackage{graphicx}
\usepackage{amsfonts}
\usepackage{epstopdf}
\usepackage{latexsym}
\usepackage{amssymb}
\usepackage{amssymb}

\usepackage[center]{subfigure}

\begin{document}

 \newcommand{\bq}{\begin{equation}}
 \newcommand{\eq}{\end{equation}}
 \newcommand{\bqn}{\begin{eqnarray}}
 \newcommand{\eqn}{\end{eqnarray}}
 \newcommand{\nb}{\nonumber}
 \newcommand{\lb}{\label}

 \preprint{YITP-18-10, IPMU18-0030}

\title{Constraints on Einstein-aether theory after GW170817}

 \author{Jacob Oost$^{a}$}
\email{Jacob_Oost@baylor.edu}

 \author{Shinji Mukohyama$^{b, c}$}
\email{shinji.mukohyama@yukawa.kyoto-u.ac.jp}

 \author{Anzhong Wang$^{a, d}$\footnote{The corresponding author}}
\email{anzhong_wang@baylor.edu}

\affiliation{$^{a}$ GCAP-CASPER, Physics Department, Baylor
University, Waco, TX 76798-7316, USA  \\
$^{b}$  Center for Gravitational Physics, Yukawa Institute for Theoretical Physics, Kyoto University, 606-8502, Kyoto, Japan \\
$^{c}$ Kavli Institute for the Physics and Mathematics of the Universe (WPI),
The University of Tokyo Institutes for Advanced Study,
The University of Tokyo, Kashiwa, Chiba 277-8583, Japan\\
$^{d}$ Institute  for Advanced Physics $\&$ Mathematics,   Zhejiang University of
Technology, Hangzhou 310032,  China }

\date{\today}

\begin{abstract}
 
 In this paper, we carry out a systematic analysis of the theoretical and observational  constraints on the dimensionless coupling constants $c_i$ ($i=1,2,3,4$) of the Einstein-aether theory, taking into account the events GW170817 and  GRB 170817A. The combination of these events restricts the deviation of the speed $c_T$ of the spin-2 graviton  to the range, $- 3\times 10^{-15} < c_T -1 < 7\times 10^{-16}$, which for the Einstein-aether theory implies $\left|c_{13}\right| \le 10^{-15}$ with $c_{ij} \equiv c_{i} + c_{j}$. The rest of the constraints are divided into two groups: those on the ($c_1, c_{14}$)-plane and those on the ($c_2, c_{14}$)-plane,  except the strong-field constraints. The latter depend on the sensitivities $\sigma_{\ae}$ of neutron stars, which are not known at present in the new ranges of the parameters found in this paper. 

 \end{abstract}


\maketitle

\section{Introduction}
 \renewcommand{\theequation}{1.\arabic{equation}} \setcounter{equation}{0}
 
 The invariance under the Lorentz symmetry group is a cornerstone of modern physics and strongly supported by experiments and observations \cite{Liberati13}. 
Nevertheless, there are various  reasons to construct gravitational theories with broken Lorentz invariance (LI) \cite{LZbreaking}. For example, if  space and/or time at the Planck scale are/is discrete, as currently  understood  \cite{QGs}, Lorentz symmetry is absent at short distance/time scales and must be an emergent low energy symmetry. A concrete example of gravitational theories with broken LI is the Ho\v{r}ava theory of quantum gravity \cite{Horava},  in which the LI  is broken via the anisotropic scaling between time and space in the ultraviolet (UV), $t \rightarrow b^{-z} t$, $x^{i} \rightarrow b^{-1} x^{i}, \; (i = 1, 2, ..., d)$, where $z$ denotes the dynamical  critical exponent, and $d$ the spatial dimensions.  Power-counting renormalizability requires $z \ge d$ at short distances, while LI demands  $z = 1$. For more details about Ho\v{r}ava gravity, see, for example, the recent review \cite{Wang17}.

Another theory that breaks LI is the Einstein-aether theory \cite{JM01}, in which LI is broken by the existence of a preferred frame defined by a time-like unit vector field, the so-called  aether field. The Einstein-aether theory is a low energy effective theory and passes all theoretical and observational constraints by properly choosing the coupling constants of the theory \cite{Jacobson}, including the stability of the Minkowski spacetime  \cite{JM04},  the abundance of the light elements formed in the early universe \cite{CL04}, gravi-\v{C}erenkov effects \cite{EMS05}, the Solar System observations \cite{FJ06}, binary pulsars \cite{Foster07a,Yagi14}, and more recently gravitational waves \cite{GHLP18}.

Among the 10 parameterized post-Newtonian (PPN) parameters \cite{Will06},  in the Einstein-aether theory the only two parameters that deviate from general relativity are $\alpha_1$ and $\alpha_2$, which measure the preferred frame effects. In terms of the four dimensionless coupling constants $c_i$'s of the Einstein-aether theory, they are given by \cite{FJ06},
\bqn
\lb{2.3aa}
\alpha_1 &=& -  \frac{8(c_3^2 + c_1c_4)}{2c_1 - c_1^2 +c_3^2 }, \nb\\
\alpha_2 &=&   \frac{1}{2}\alpha_1  - \frac{(c_1 +2c_3 - c_4)(2c_1 + 3c_2+c_3+c_4)}{c_{123}(2-c_{14})}, ~~~~~~~~
\eqn
where $c_{ij} \equiv c_i + c_j$ and $c_{ijk} = c_i + c_j + c_k$. In the weak-field regime, using lunar laser ranging and solar alignment with the ecliptic, 
Solar System observations constrain these parameters to very small values \cite{Will06},
\bq
\lb{CD5}
\left| \alpha_1\right| \le 10^{-4}, \quad 
 \left|\alpha_2\right| \le 10^{-7}.
 \eq

 Considering the smallness of $\alpha_A$ ($A=1,2$), it may be convenient to Taylor expand Eq.(\ref{2.3aa}) with respect to $\alpha_A$ to obtain
\bq
 \lb{eq3.10}
c_2 = - \frac{c_{13}(2c_1 - c_3)}{3c_1} + \mathcal{O}(\alpha_A)\,, \quad
c_4 =  - \frac{c_3^2}{c_1} + \mathcal{O}(\alpha_A)\,.
\eq
If terms of order $\mathcal{O}(\alpha_A)$ and higher are small enough to be neglected then the four-dimensional parameter space spanned by $c_i$'s reduces to two-dimensional one. Until recently, the strongest constraints on the Einstein-aether theory were (\ref{CD5}) and thus this treatment was a good approximation. Then, using the order-of-magnitude arguments about the orbital decay of binary pulsars, Foster estimated that $|c_1\pm c_3| \lesssim {\cal{O}}\left(10^{-2}\right)$, by further assuming that $c_i \ll 1$ \cite{Foster07a}. More detailed analysis of binary pulsars showed that $c_{13} \lesssim {\cal{O}}\left(10^{-2}\right)$, $|c_1-c_3| \lesssim {\cal{O}}\left(10^{-3}\right)$ (See Fig. 1 in \cite{Yagi14}).

However, the combination of the gravitational wave event GW170817 \cite{GW170817}, observed by the LIGO/Virgo collaboration, and the one of the gamma-ray burst GRB 170817A \cite{GRB170817}, provides  much more severe constraint on $c_{13}$. In fact,  these events imply that the speed of the spin-2 mode $c_T$ must satisfy the bound, $- 3\times 10^{-15} < c_T -1 < 7\times 10^{-16}$. In the Einstein-aether theory, the speed of the spin-2 graviton is given by $c_{T}^2 = 1/(1-c_{13})$ \cite{JM04}, so the  GW170817 and GRB 170817A events imply  
\bq
\lb{CDa} 
\left |c_{13}\right| < 10^{-15}.
 \eq
 This is much smaller than the limits of Eq.(\ref{CD5}). As a result, if we still adopt the Taylor expansion with respect to $\alpha_A$ then Eq.(\ref{eq3.10}), for example, can  no longer  be approximated only up to the zeroth-order of $\alpha_A$. Instead, it must be expanded  at least up to the fourth-order of $\alpha_1$, the second-order of $\alpha_2$ (plus their mixed terms), and the first-order of $c_{13}$, in order to obtain a consistent treatment. Otherwise, the resulting errors would become much larger than $\left|c_{13}\right|$, due to the omissions of the terms higher in $\alpha_{A}$,  and the results obtained in this way would not be trustable.

 In this paper, we shall therefore Taylor expand all constraints other than (\ref{CDa}) with respect to $c_{13}$, keep only terms zeroth order in $c_{13}$ by setting $c_{13}\simeq 0$ in those expressions, and let $c_1$, $c_2$ and $c_{14}$ be restricted by those other constraints. (In particular, we shall not set $\alpha_A \simeq 0$ since this would cause large errors.) As a result, the phase space of $c_i$'s becomes essentially three-dimensional. Moreover, it is to our surprise that the three-dimensional phase space actually becomes degenerate, in the sense that the constraints can be divided into two groups, one has constraints only on the ($c_1, c_{14}$)-plane, and the other has  constraints only on the ($c_2, c_{14}$)-plane \footnote{Note that in \cite{Hansen15} the case $c_{13} = \alpha_2 = 0$ was considered, so the parameter space was again reduced to two-dimensional. Then, the constraints  were restricted to  the $\left(\alpha_1, c_-\right)$-plane, where $c_- \equiv c_1 - c_3$. It was found that in this case no bounds can be imposed on $c_{-}$.}.

The rest of the paper is organized as follows: In Sec. II we briefly review the  Einstein-aether theory. In Sec. III we first list all the relevant constraints, theoretical and observational, then consider them one by one, and finally obtain a region in the phase space, in which all theoretical and observational constraints are satisfied by the Einstein-aether theory, except for the strong-field constraints given by Eq.(\ref{CD6}). These strong-field constraints depend on the sensitivities $\sigma_{\ae}$ of neutron stars in the Einstein-aether theory, which depends on $c_i$'s (and the equation of state of nuclear matter) \cite{Yagi14} and are not known for the new ranges of the parameters found in this paper. Thus, we shall not use these strong-field constraints to obtain further constraints on $c_i$'s, leaving further studies to a future work. Our main results are summarized in Sec. IV, in which   some concluding remarks and discussion are also presented.

\section{Einstein-Aether Theory}
 \renewcommand{\theequation}{2.\arabic{equation}} \setcounter{equation}{0}

In  Einstein-aether ($\ae$-) theory, the fundamental variables of the gravitational  sector are \cite{JM01}, 
\bq
\lb{2.0a}
\left(g_{\mu\nu}, u^{\mu}, \lambda\right),
\eq
with the Greek indices $\mu,\nu = 0, 1, 2, 3$, and $g_{\mu\nu}$ is  the four-dimensional metric  of the space-time
with the signatures $(-, +,+,+)$,  $u^{\mu}$ the aether four-velocity, and $\lambda$ is a Lagrangian multiplier, which guarantees that the aether  four-velocity  is always timelike. 
The general action of the theory  is given by  \cite{Jacobson},
\bq
\lb{2.0}
S = S_{\ae} + S_{m},
\eq
where  $S_{m}$ denotes the action of matter,  and $S_{\ae}$  the gravitational action of the $\ae$-theory, given by
\bqn
\lb{2.1} 
 S_{\ae} &=& \frac{1}{16\pi G_{\ae}}\int{\sqrt{- g} \; d^4x \Big[R(g_{\mu\nu}) + {\cal{L}}_{\ae}\left(g_{\mu\nu}, u^{\lambda}\right)\Big]},\nb\\  
S_{m} &=& \int{\sqrt{- g} \; d^4x \Big[{\cal{L}}_{m}\left(g_{\mu\nu}, \psi\right)\Big]}.
\eqn
Here $\psi$ collectively denotes the matter fields, $R$    and $g$ are, respectively, the  Ricci scalar and determinant of $g_{\mu\nu}$, 
 and 
\bq
\lb{2.2}
 {\cal{L}}_{\ae}  \equiv - M^{\alpha\beta}_{~~~~\mu\nu}\left(D_{\alpha}u^{\mu}\right) \left(D_{\beta}u^{\nu}\right) + \lambda \left(g_{\alpha\beta} u^{\alpha}u^{\beta} + 1\right),
\eq
 where $D_{\mu}$ denotes the covariant derivative with respect to $g_{\mu\nu}$, and  $M^{\alpha\beta}_{~~~~\mu\nu}$ is defined as
\bqn
\lb{2.3}
M^{\alpha\beta}_{~~~~\mu\nu} = c_1 g^{\alpha\beta} g_{\mu\nu} + c_2 \delta^{\alpha}_{\mu}\delta^{\beta}_{\nu} +  c_3 \delta^{\alpha}_{\nu}\delta^{\beta}_{\mu} - c_4 u^{\alpha}u^{\beta} g_{\mu\nu}.\nb\\
\eqn
Note that here we assume that matter fields couple only to $g_{\mu\nu}$, so ${\cal{L}}_{m}$ is independent of $u^{\mu}$.

The four coupling constants $c_i$'s are all dimensionless, and $G_{\ae}$ is related to  the Newtonian constant $G_{N}$ via the relation \cite{CL04},
\bq
\lb{2.3a}
G_{N} = \frac{G_{\ae}}{1 - \frac{1}{2}c_{14}}.
\eq

The variations of the total action with respect to $g_{\mu\nu},\; u^{\mu}$ and $\lambda$ yield, respectively, the field equations,
 \bqn
 \lb{2.4a}
 E^{\mu\nu} &=& 8\pi G_{\ae} T_{m}^{\mu\nu},\\
 \lb{2.4b}
  \AE_{\mu} &=& 0, \\
   \lb{2.4c}
  g_{\alpha\beta} u^{\alpha}u^{\beta} &=& -1, 
 \eqn
where
 \bqn
 \lb{2.5}
 E^{\mu\nu} &\equiv& R^{\mu\nu} - \frac{1}{2}g_{\mu\nu}R - T^{\mu\nu}_{\ae},\nb\\
 T_{m}^{\mu\nu} &\equiv&  \frac{2}{\sqrt{-g}}\frac{\delta \left(\sqrt{-g} {\cal{L}}_{m}\right)}{\delta g_{\mu\nu}},\nb\\
  T_{\ae}^{\alpha\beta} &\equiv&  
  -D_{\mu}\Big[u^{(\beta}J^{\alpha) \mu} - J^{\mu(\alpha}u^{\beta)} - J^{(\alpha\beta)}u^{\mu}\Big]\nb\\
&& - c_1\Big[\left(D_{\mu}u^{\alpha}\right)\left(D^{\mu}u^{\beta}\right) - \left(D^{\alpha}u_{\mu}\right)\left(D^{\beta}u^{\mu}\right)\Big]\nb\\
&& + c_4 a^{\alpha}a^{\beta}    + \lambda  u^{\alpha}u^{\beta} - \frac{1}{2}  g^{\alpha\beta} J^{\delta}_{\;\;\sigma} D_{\delta}u^{\sigma},\nb\\
 \AE_{\mu} & \equiv &
 D_{\alpha} J^{\alpha}_{~~~\mu} + c_4 a_{\alpha} D_{\mu}u^{\alpha} + \lambda u_{\mu}, 
 \eqn
 with
\begin{equation}
 \lb{2.6}
J^{\alpha}_{\;\;\;\mu} \equiv M^{\alpha\beta}_{~~~~\mu\nu}D_{\beta}u^{\nu}\,,\quad
a^{\mu} \equiv u^{\alpha}D_{\alpha}u^{\mu}\,.
\end{equation}
From Eqs.(\ref{2.4b}) and (\ref{2.4c}),  we find that
\bq
\lb{2.7}
\lambda = u_{\beta}D_{\alpha}J^{\alpha\beta} + c_4 a^2\,,
\eq
where $a^{2}\equiv a_{\lambda}a^{\lambda}$.

 \section{Constraints after GW170817}
 \renewcommand{\theequation}{3.\arabic{equation}} \setcounter{equation}{0}

 It is easy to show that the Minkowski spacetime is a solution of the Einstein-aether theory, in which the aether is aligned along the time direction, $\bar{u}_{\mu} = \delta^{0}_{\mu}$. It is then straightforward to analyze linear perturbations around the Minkowski background and investigate properties of spin-$0$, -$1$ and -$2$ excitations (see Appendix A  and/or ref.~\cite{Foster06a} for details). In particular, the coefficients of the time kinetic term of each excitation $q_{S,V,T}$ must be positive~\footnote{\label{footnote:decouplinglimit}In the so-called decoupling limit $c_i\to 0$, $q_V=c_{14}$ vanishes but the limit must be taken from the positive side of $q_{S,V,T}$ and $c_{S,V,T}^2$. Similarly, if we would like to take the infinite speed limit, e.g. $c_{S}\to\infty$, it should also be taken from the positive side.}:
\begin{equation}
\lb{qsvt}
q_{S,V,T} > 0\,, 
\end{equation}
where
\begin{eqnarray}
 q_S & = & \frac{\left(1-c_{13}\right)\left(2+ c_{13} + 3c_2\right)}{c_{123}}\,,\nonumber\\
 q_V & = & c_{14}\,,\nonumber\\
 q_T & = & 1-c_{13}\,.
\end{eqnarray}
In addition to the ghost-free condition for each part of the linear perturbations, we must also require the theory be free of gradient instability, that is,  the squared speeds must be non-negative,
\bq
\lb{CD1}
c_{S,V,T}^2 \ge 0\,,
\eq
where 
\begin{eqnarray}
 \label{eqn:soundspeeds}
 c_S^2 & = & \frac{c_{123}(2-c_{14})}{c_{14}(1-c_{13}) (2+c_{13} + 3c_2)}\,,\nonumber\\
 c_V^2 & = & \frac{2c_1 -c_{13} (2c_1-c_{13})}{2c_{14}(1-c_{13})}\,,\nonumber\\
 c_T^2 & = & \frac{1}{1-c_{13}}\,.
\end{eqnarray}
Moreover, $c_{S,V,T}^2-1$ must be greater than $-10^{-15}$ or so, in order to avoid the existence of the vacuum gravi-\v{C}erenkov radiation by matter such as cosmic rays \cite{EMS05}. We thus impose
 \bq
\lb{CD2}
c_{S,V,T}^2 \gtrsim 1\,, 
\eq
which is stronger than (\ref{CD1}).

More recently, as mentioned above,  the combination of the gravitational wave event GW170817 \cite{GW170817}, observed by the LIGO/Virgo collaboration, and the event of the gamma-ray burst GRB 170817A \cite{GRB170817} provides  a remarkably stringent constraint on the speed of the spin-2 mode, $- 3\times 10^{-15} < c_T -1 < 7\times 10^{-16}$, which implies the constraint (\ref{CDa}).

On the other hand, applying the theory to cosmology, it was found that the gravitational constant appearing in the effective Friedman equation is given by \cite{CL04},
\bq
\lb{2.3ga}
G_{{\mbox{cos}}} = \frac{G_{\ae}}{1+\frac{1}{2}(c_{13} + 3c_2)}.  
 \eq
Since $G_{{\mbox{cos}}}$ is not the same as $G_N$ in (\ref{2.3a}), the expansion rate of the universe differs from what would
have been expected in GR.     
In particular, decreasing 
  the Hubble expansion rate  during the big bang nucleosynthesis  will result in weak interactions freezing-out later, and leads to a lower freeze-out temperature. 
This will yield a decrease in the production of the primordial $^{4}$He, and subsequently a lower  $^{4}$He-to-hydrogen mass ratio  \cite{CL04}.
 As a result the primordial helium abundance is modified, and to be consistent with current observations \cite{COb}, 
the ratio must satisfy the  constraint,
\bq
\lb{CD4}
\left|\frac{G_{{\mbox{cos}}}}{G_N} - 1\right| \lesssim \frac{1}{8}.
 \eq
 One could obtain other cosmological constraints on $G_{{\mbox{cos}}}/G_N$ if we make assumptions on the dark sector of the universe \cite{Frusciante:2015maa}. While they are interesting and important, we shall not consider those additional constraints  since they are model-dependent.

Moreover, for any choice of $c_i$'s, all PPN parameters \cite{Will06}  of the $\ae$-theory agree with those of GR \cite{EJ04,FJ06},  except
the preferred frame parameters which are given by Eq.(\ref{2.3aa})  \cite{FJ06,F06,GJW05}. 
In the weak-field regime, using lunar laser ranging and solar alignment with the ecliptic, Solar System observations constrain these parameters to very small values (\ref{CD5}) \cite{Will06}.
In the strong-field regime, using data from the isolated millisecond pulsars PSR B1937 + 21 \cite{SPulsarA} and PSR J17441134  \cite{SPulsarB},  the following constraints were obtained \cite{SW13},
\bq
\lb{CD6}
\left|\hat\alpha_1\right| \le 10^{-5}, \quad 
 \left|\hat\alpha_2\right| \le 10^{-9},
 \eq
at $95\%$ confidence, where ($\hat\alpha_1, \hat\alpha_2$) denotes the strong-field generalization of  ($\alpha_1, \alpha_2$)  \cite{DEF92}. 
In the Einstein-$\ae$ther theory, they are given by \cite{Yagi14}, 
\bqn
\lb{2.3ac}
\hat\alpha_1 &=& \alpha_1 + \frac{c_-(8+\alpha_1)\sigma_{\ae}}{2c_1}, \nb\\
\hat\alpha_2&=& \alpha_2 + \frac{\hat\alpha_1 - \alpha_1}{2}   
- \frac{(c_{14} -  2)(\alpha_1 -2\alpha_2)\sigma_{\ae}}{2(c_{14} - 2c_{13})},~~~~~
\eqn
where 
$\sigma_{\ae}$ denotes the sensitivity.

To consider the above constraints, one may first express two of the four parameter $c_n$'s, say, $c_2$ and $c_4$,  in terms of $\alpha_A$'s   through Eqs.(\ref{2.3aa}), and then expand $c_2$ and $c_4$ in terms of $\alpha_A$, as given by Eq.(\ref{eq3.10}).  Thus, to the zeroth-order of $\alpha_A$'s, $c_2$ and $c_4$ are given by the first term in each of Eq.(\ref{eq3.10})  \cite{FJ06,Jacobson}.
In fact, this is what have been doing so far in the analysis of the observational constraints of the Einstein-aether theory \cite{Jacobson,Foster06a,Yagi14,GHLP18}.

However, with the new constraint (\ref{CDa}), if we still adopt the Taylor expansion with respect to $\alpha_A$,  then, to have a self-consistent expansion, one must expand $c_2$ and $c_4$ at least up to the fourth-order of $\alpha_1$, the second-order of $\alpha_2$ (plus their mixed terms, such as $\alpha_1^2\alpha_2$) [cf. Eq.(\ref{CD5})], and the first-order of $c_{13}$.  Clearly, this will lead to very complicated analyses. In the following, instead, we simply Taylor expand constraints other than (\ref{CDa}) with respect to $c_{13}$, keep only terms zeroth order in $c_{13}$, and let all the other parameters constrained by those approximated constraints. Then, keeping  only the leading terms in the $c_{13}$-expansion is equivalent to setting
\bq
\lb{3.1} 
 c_{13} = 0\,.
 \eq
As a result,  the errors are of the order of ${\cal{O}}\left(10^{-15}\right)$, as far as Eq.(\ref{CDa}) is concerned. Thus,  the resulting errors due to this omission is insignificant, in comparison to the bounds of  the rest of the observational constraints.
 Hence, while the constraint $q_T>0$ is automatically satisfied, $q_S>0$ yields
\bqn
\lb{cd1a} 
  \frac{2+ 3c_2}{c_{2}} > 0\,.
 \eqn

On the other hand,  from Eqs.(\ref{eqn:soundspeeds}) and (\ref{2.3aa}) we find that 
\bqn
\lb{3.2} 
c_V^2 = \frac{c_1}{c_{14}}\,, \quad \alpha_1 = - 4  c_{14}\,,
 \eqn
so the constraints (\ref{CD5}), $q_V>0$ and $c_V^2\gtrsim 1$ lead to
\bqn
\lb{3.3} 
0 < c_{14} \le 2.5\times 10^{-5}, \quad 
c_{14} \lesssim c_1\,.
\eqn

It is remarkable that these two constraints are all confined to the ($c_1, c_{14}$)-plane,  while the rest are  all confined to the ($c_2, c_{14}$)-plane, as to be shown below. As we shall see, this considerably simplifies the analysis of the whole set of the constraints listed above.

In particular,  the constraint (\ref{CD4}) is reduced to 
\bq
\lb{3.4}
- \frac{1}{8} \lesssim \frac{c_{14} + 3c_2}{2 + 3c_2} \lesssim \frac{1}{8}\,,
\eq
which is rewritten as
\begin{equation}
 - \frac{2(1+4c_{14})}{27} \lesssim c_2 \lesssim \frac{2(1-4c_{14})}{21}\,.
\end{equation}
Considering the fact that $|c_{14}|$ is as small as (\ref{3.3}),  we then find that 
\bqn
\lb{3.8}
- \frac{2}{27} \lesssim c_2 \lesssim  \frac{2}{21}\,,
\eqn
which, together with the constraint (\ref{cd1a}), yields, 
\bqn
\lb{3.8a}
 0 < c_2 \lesssim  0.095.
\eqn

On the other hand, from $c_S^2 \gtrsim 1$ we also find that 
\bq
\lb{3.9}
\frac{c_2(2-c_{14})}{c_{14}(2 + 3c_2)} \gtrsim 1\,.
\eq
Considering the constraints (\ref{3.3}) and (\ref{3.8a}), we find that Eq.(\ref{3.9})  is equivalent to  
\bq
\lb{3.10}
0 <  c_{14} \lesssim c_2\,,
\eq
which,  together with the constraint (\ref{3.8a}),  yields
\bqn
\label{eqn:c2range}
 0 < c_{14} \lesssim c_2 \lesssim  0.095\,.
\eqn

\begin{figure}
{
\includegraphics[width=8cm]{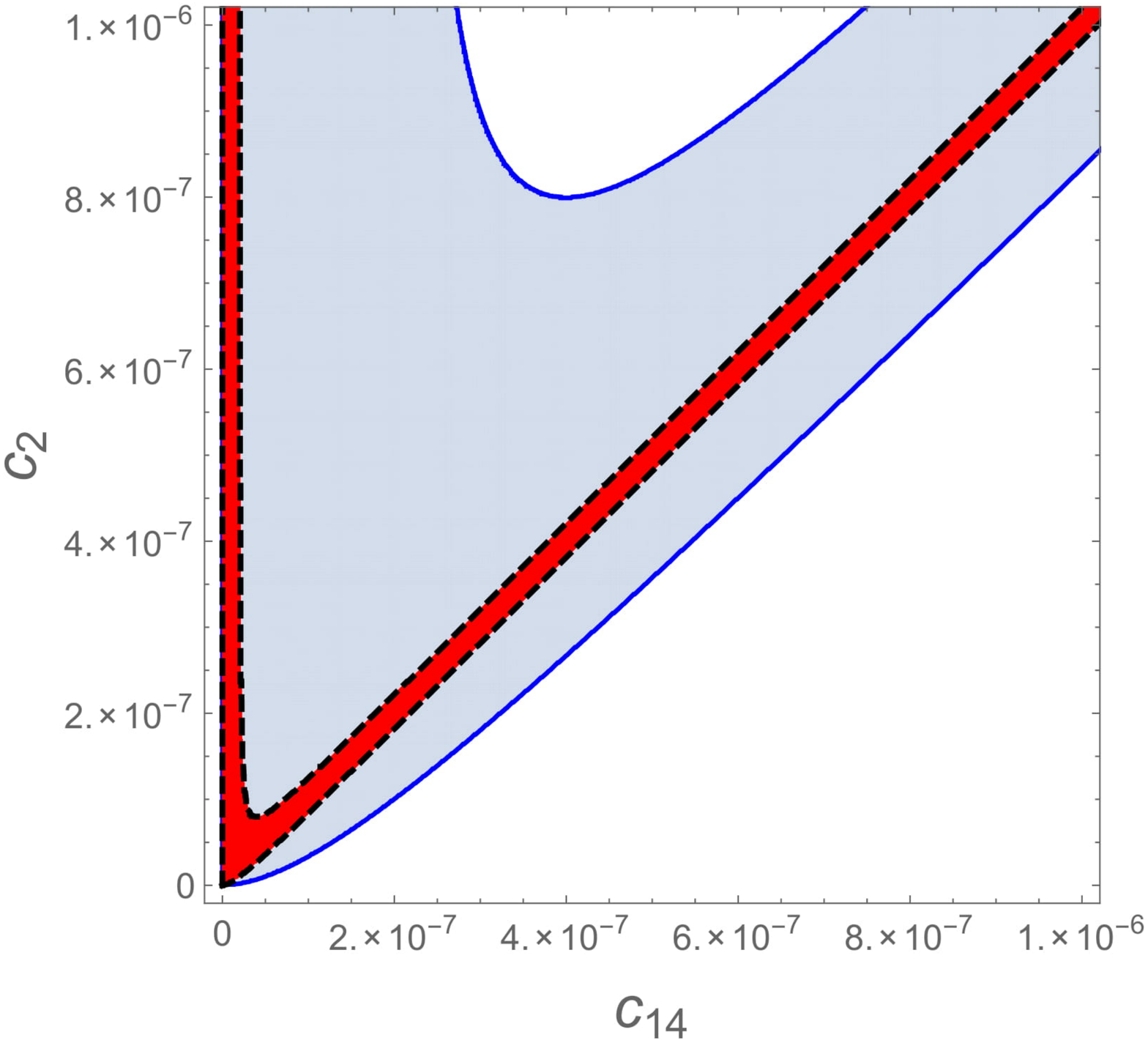}
\includegraphics[width=8cm]{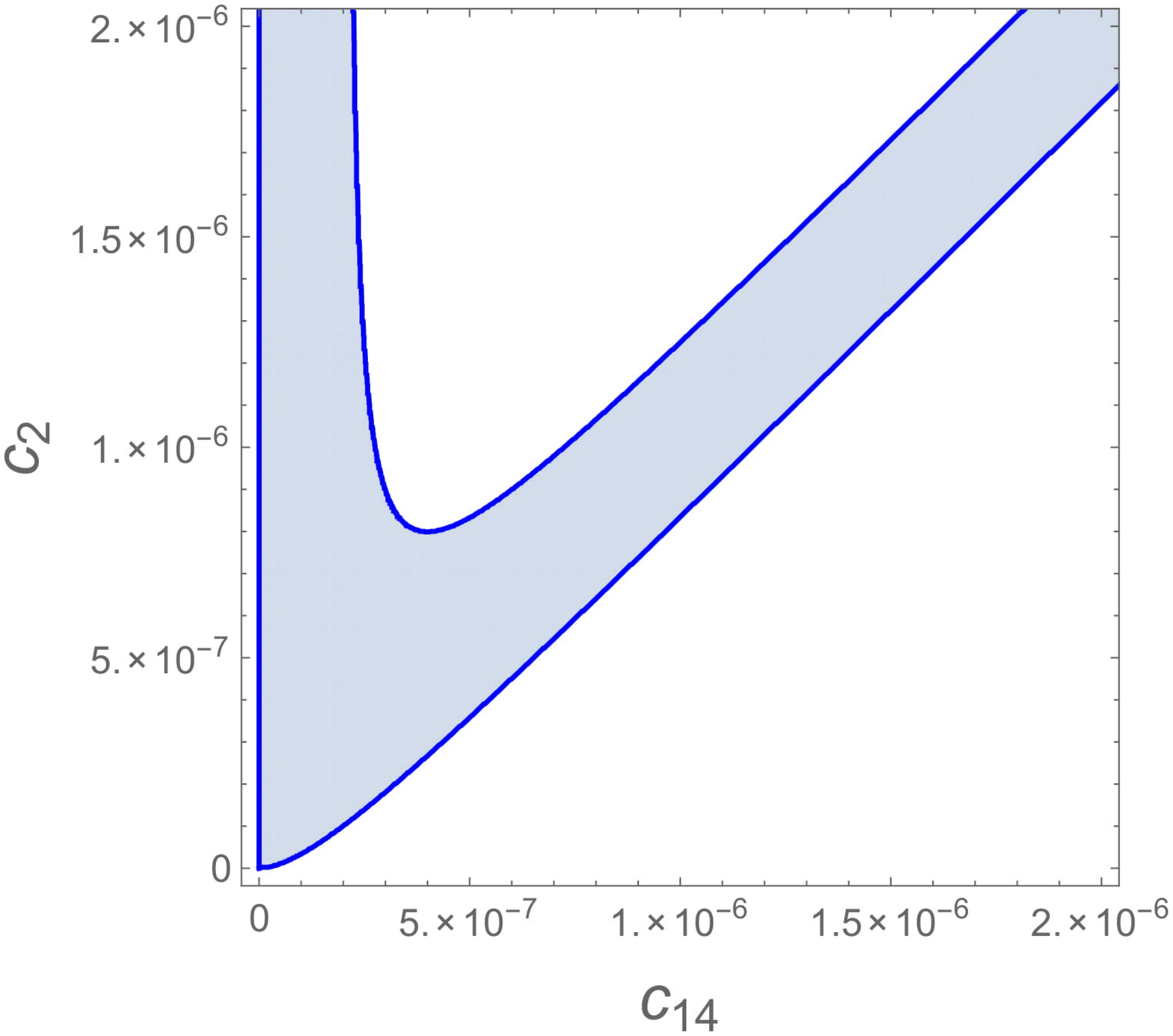}
\includegraphics[width=8cm]{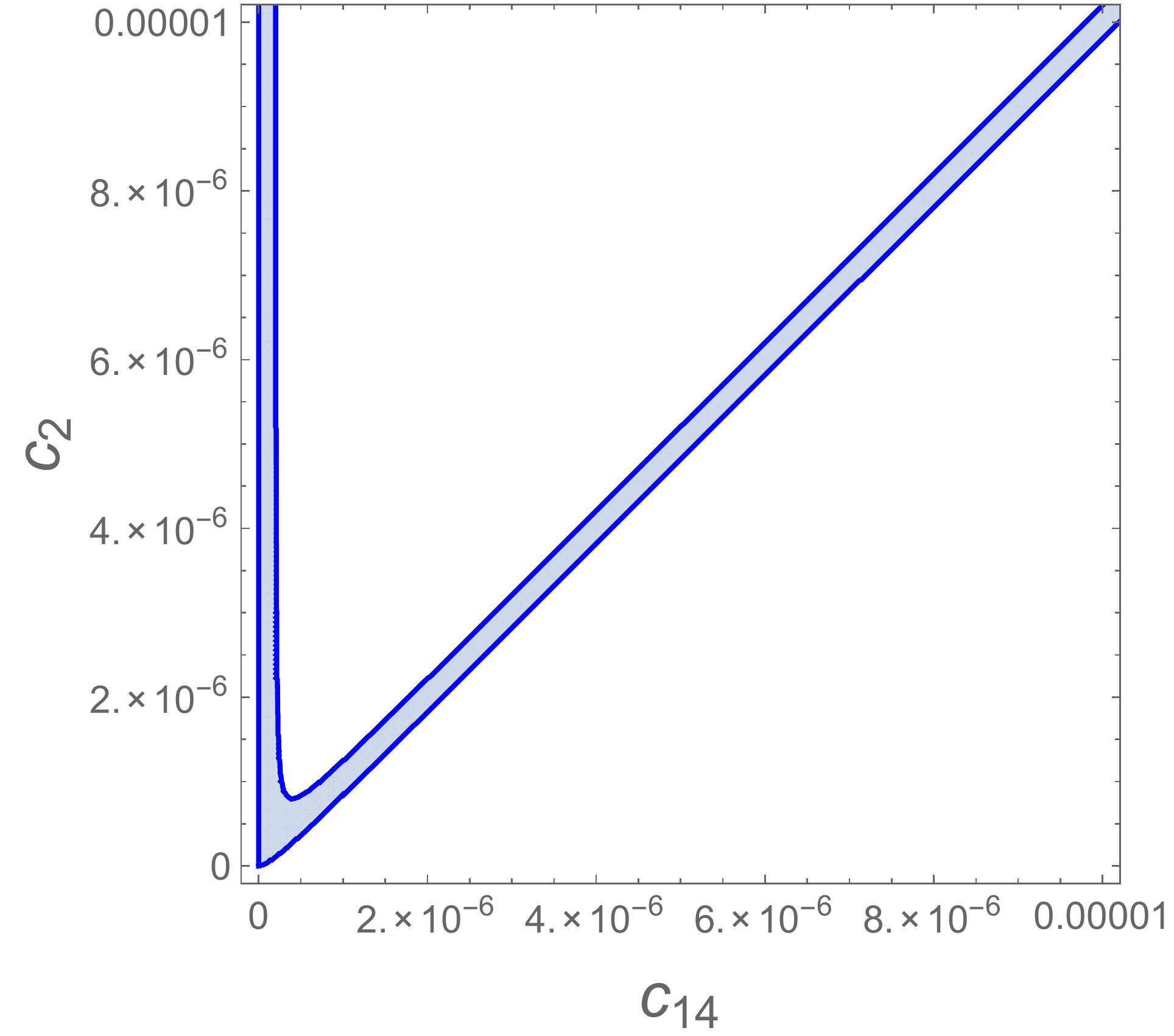}
}
\caption{In this figure, we plot the   constraint $\left|\alpha_2\right| \le 10^{-7}$ given by Eq.(\ref{3.12}), together with (\ref{eqn:c2range}), in the ($c_2, c_{14}$)-plane. In each of the plots different scales of the horizontal axis have been chosen. In the top plot, the region $\left|\alpha_2\right| \le 10^{-8}$ marked with red color and dashed line boundary  is also shown. }
\label{fig1}
\end{figure}

By setting $c_{13} = 0$ in Eq.(\ref{2.3aa}), we also find
\bq
\lb{3.11}
\alpha_2 \simeq \frac{c_{14}\left(c_{14} + 2c_2c_{14} - c_2\right)}{c_2\left(2-c_{14}\right)}, 
\eq
and the second constraint in (\ref{CD5}) yields
\bq
\lb{3.12}
- 10^{-7} \le  \frac{c_{14}\left(c_{14} + 2c_2c_{14} - c_2\right)}{c_2\left(2-c_{14}\right)} \le 10^{-7}. 
\eq
In Fig.~\ref{fig1}, we show this constraint, combined with (\ref{eqn:c2range}), for various scales of $c_{14}$ in the ($c_2, c_{14}$)-plane. The constraints in the ($c_2, c_{14}$)-plane have simple expressions for values of $c_{14}$ smaller than $2\times 10^{-7}$ or sufficiently larger than $2\times 10^{-7}$ (say, for $c_{14}$ larger than $2\times 10^{-6}$): the constraints are satisfied in either of the following two regions,
\begin{eqnarray}
\lb{CSb}
 \mbox{(i)}  & & 
  0<c_{14}\leq 2\times 10^{-7}\,, \nonumber\\
 & &  c_{14} \lesssim c_2 \lesssim  0.095\,, \nonumber\\
 \mbox{(ii)} &&
  2\times 10^{-6}\lesssim c_{14}\lesssim 2.5\times 10^{-5}\,,\nonumber\\
 & &  0 \lesssim c_2-c_{14} \lesssim 2\times 10^{-7}\,.
\end{eqnarray}
For the constraints in the intermediate regime of $c_{14}$ ($2\times 10^{-7}< c_{14}\lesssim 2\times 10^{-6}$), see the top and the middle plots in Fig.~\ref{fig1}.

The constraints (\ref{CD6}) with (\ref{2.3ac}) in principle constrain the parameters $c_i$'s. However, the sensitivities $\sigma_{\ae}$ of a neutron star, which depend on $c_i$'s and the equation of state of nuclear matter \cite{Yagi14}, are  not known so far within the new ranges of the parameters given above. Therefore, instead of using (\ref{CD6}) to constrain the parameters $c_i$'s, we simply rewrite them in term  of $c_i$'s and the sensitivities $\sigma_{\ae}$ for future references. Setting  $c_{13} = 0$ in Eq.(\ref{2.3ac}), we find that
 \bqn
\lb{3.13}
\hat\alpha_1 &=&
\alpha_1\left[ 1 + \sigma_{\ae}  \left(1 + \frac{8}{\alpha_1}\right)\right]\,,\nonumber\\
\hat\alpha_2&=&
\alpha_2\left[ 1 + \sigma_{\ae}  \left(1 + \frac{8}{\alpha_1}\right)\right]\,.
\eqn
Since $\left|\alpha_1\right| \le 10^{-4}$, the constraints (\ref{CD6}) are reduced to 
\begin{equation}
 \lb{3.14}
  \left|\alpha_1 + 8\sigma_{\ae}\right| \le 10^{-5}\,,  \quad
  \left|\frac{\alpha_2}{\alpha_1}\right| \times
  \left|\alpha_1 + 8\sigma_{\ae}\right|  \le 10^{-9}\,. 
\end{equation}
As already mentioned above, we leave the analysis of these two constraints that involves the computation of the sensitivities $\sigma_{\ae}$ to a future work.

\section{Discussions  and Conclusions}
 \renewcommand{\theequation}{4.\arabic{equation}} \setcounter{equation}{0}

In this paper, we have considered various constraints on the Einstein-aether theory, as listed in Eqs.(\ref{CDa}), (\ref{qsvt})-(\ref{CD6}), which represent the major constraints from the self-consistency of the theory to various observations. The severest one is from the recent  gravitational wave event,  GW170817 \cite{GW170817}, observed by the LIGO/Virgo collaboration, and  the gamma-ray burst observation of GRB 170817A \cite{GRB170817}, given by Eq.(\ref{CDa}) due to the constraint on the deviation of the speed of the spin-2 graviton from that of light. 

In the previous studies, all analyses were done by expanding the two parameters $c_2$ and $c_4$ in terms of $\alpha_1$ and $\alpha_2$ through the relations given by Eq.(\ref{eq3.10}), 
and then keeping only the leading terms, so finally  one obtains \cite{FJ06,Jacobson},
\bqn
\lb{4.1}
c_2 = - \frac{c_{13}(2c_1 - c_3)}{3c_1}, \quad c_4 =  - \frac{c_3^2}{c_1}, \; (\alpha_1 = \alpha_2 = 0). ~~~
\eqn
Clearly, in this approach  the errors due to the omission of the higher-order terms are of the order of ${\cal{O}}(\alpha_1) \simeq 10^{-4}$, which is too large in comparing with the new constraint  
(\ref{CDa}) from the observations  of gravitational waves \cite{GW170817,GRB170817}.

In this paper, instead, for any given constraint, say, $F(c_i) = 0$,  we have expanded it only in terms of $\epsilon \equiv c_{13}$, 
\bqn
\lb{4.2}
&& F(c_1, c_2, c_{14}, \epsilon) =   F(c_1, c_2, c_{14}, 0) \nb\\
&& ~~~~~~~ +  F_{,\epsilon}(c_1, c_2, c_{14}, 0)\epsilon + ... = 0,
\eqn
and leave all the other parameters free.  Then, keeping only the leading term,  we can  see that the resulting  errors due to this omission is of the order of ${\cal{O}}\left(10^{-15}\right)$, which is insignificant
 in comparing with the rest of constraints.   In doing so, the reduced phase space is in general three-dimensional. However, it is remarkable that the   constraints are then divided into two groups,
one is confined on  the ($c_1, c_{14}$)-plane, and the other on   the ($c_2, c_{14}$)-plane. In the former, the constraints are 
given by Eq.(\ref{3.3}). We can also transfer  this constraint to the ($c_4, c_{14}$)-plane, which is simply equal to,
\bqn
\lb{4.3}
c_4 \lesssim 0, \quad0 < c_{14} \le 0.25\times 10^{-4}. 
\eqn
(See footnote~\ref{footnote:decouplinglimit} for a comment on the $c_{14}\to 0$ limit.)

On the other hand, the cosmological constraint from the measurements of the primordial helium-4  abundance restricts  $c_2$ to the range given by Eq.(\ref{3.8a}),  while the constraint $c_S^2 \gtrsim 1$ further requires,
\bq
\lb{4.4}
0.095 \gtrsim c_2 \gtrsim c_{14}  > 0.
\eq
 (see footnote~\ref{footnote:decouplinglimit} again). However,  the severest constraint on $c_2$ comes from Eq.(\ref{3.12}), from which we find the constraints (\ref{CSb}) for $c_{14} \in \left[0,  \; 2\times 10^{-7}\right]$ and $  c_{14} \in \left[2\times 10^{-6}, \;  2.5\times 10^{-5}\right]$, respectively. In the intermediate regime, $c_{14} \in \left(2\times 10^{-7}, \;  2\times 10^{-6}\right)$, the constraints are illustrated in the top and  middle plots in Fig.~\ref{fig1}.

It should be noted that the constraints given above do not include the strong-field regime constraints (\ref{CD6}), because they depend on the sensitivities of neutron stars in the theory, which are not known 
so far for the parameters given in the above new ranges   \cite{Yagi14}. Therefore, instead using them to put further constraints on the parameter $c_i$'s,  we have used them to find the upper bounds on  the sensitivity 
parameter $\sigma_{\ae}$,  given by Eq.(\ref{3.14}), i.e.,
\begin{equation}
 \lb{4.5}
  \left|\alpha_1 + 8\sigma_{\ae}\right| \le 10^{-5}\,,  \quad
  \left|\frac{\alpha_2}{\alpha_1}\right| \times
  \left|\alpha_1 + 8\sigma_{\ae}\right|  \le 10^{-9}\,, 
\end{equation}
although  they are not free parameters, and normally depend on $c_i$'s, as shown explicitly in \cite{Yagi14}. Eq.(\ref{4.5}) represents  very severe constraints, and imposes tight bounds 
 on the radiation of neutron stars in the Einstein-aether theory, through  the emissions of the different species of the spin-0, spin-1 and spin-2 gravitons. Therefore, it would be very interesting to calculate 
 $\sigma_{\ae}$ in the new ranges of the free parameters $c_i$'s, and then comparing such obtained
values of $\sigma_{\ae}$  with the constraints (\ref{4.5}). 

Finally, we note that recently  constraints of the khronometric  theory \cite{BPS11} was studied numerically  in \cite{GSS18}. When the aether is hypersurface-orthogonal, 
\bq
\lb{4.6}
u_{[\alpha}D_{\beta}u_{\lambda]} = 0,
\eq
it can be shown that   $u_{\mu}$ can  be always written in terms of a timelike scalar field $\phi$, {\em  the khronon}, in the form \cite{Wald94}, 
\bq
\lb{4.7}
u_{\mu}= \frac{\phi_{,\mu}}{\sqrt{-\phi_{,\alpha}\phi^{,\alpha}}}, \quad \phi_{,\alpha}\phi^{,\alpha} < 0.
\eq
Then, we find that, 
\bq
\lb{4.8}
\omega^2 \equiv  a^{\mu}a_{\mu} + \big(D_{\alpha}u_{\beta}\big)\big(D^{\alpha}u^{\beta}\big) -   \big(D_{\alpha}u_{\beta}\big)\big(D^{\beta}u^{\alpha}\big),
\eq
vanishes identically. As a result, one can add the following term to the general action (\ref{2.0})  \cite{J10,Wang12},
\bq
\lb{4.9}
\Delta S_{\ae}   \equiv c_{0} \int{dx^4\sqrt{-g} \;  \omega^2}, 
\eq
where $c_{0}$ is an arbitrary dimensionless constant. Hence, among the four coupling constants $c_i\; (i = 1,  3, 4)$ of the Einstein-aether theory, only the three combinations ($c_{14}$, $c_{13}$, $c_2$) have physical meaning in the khronometric  theory \cite{BPS11}. This theory was also referred to as the ``T-theory'' in \cite{J10} \footnote{It is interesting to note that the khronometric  theory can be considered as the low energy limit of the non-projectable version of the Ho\v{r}ava gravity \cite{J10,BPS11,Wang12,Wang17}.}.

In view of the above considerations, it is clear that the spin-1 graviton appearing in the Einstein-aether theory is absent in the khronometric  theory \footnote{In addition, an instantaneous mode appears in the khronometric theory \cite{BS11,Wang17,LMWZ17}, while this mode is absent in the Einstein-aether theory \cite{Jacobson,JM04}.}. As a result, all the constraints from the spin-1 mode should be dropped, in order to obtain the constraints on the khronometric  theory. In other words, the constraints obtained in the present paper projected onto the three dimensional subspace ($c_{14}$, $c_{13}$, $c_2$) are more stringent than the constraints found in \cite{GSS18}.

\section*{\bf Acknowledgements}
 
We would like to thank K. Lin for valuable comments. The work of A.W. was supported in part by   the National Natural Science Foundation of China (NNSFC), Grant Nos. 11375153 and 11675145.  
The work of S.M. was supported by Japan Society for the Promotion of Science (JSPS) Grants-in-Aid for Scientific Research (KAKENHI) No. 17H02890, No. 17H06359, and 
by World Premier International Research Center Initiative (WPI), MEXT, Japan.

\vspace{1cm}

\section*{Appendix A: Linear perturbations around Minkowski background}
\label{app:linearperturbation}
 \renewcommand{\theequation}{A.\arabic{equation}} \setcounter{equation}{0}

 It is easy to show that the Minkowski spacetime is a solution of the Einstein-aether theory, in which the aether is aligned along the time direction,
$\bar{u}_{\mu} = \delta^{0}_{\mu}$. Let us consider the linear perturbations,
\bq
\lb{eq3.1}
g_{\mu\nu} = \eta_{\mu\nu} + h_{\mu\nu}, \quad u_{\mu} = \bar{u}_{\mu} + w_{\mu},
\eq
where 
\bqn
\lb{eq3.2}
 h_{0i} &=& \partial_i B + B_i\,, \quad w_{i} = \partial_i v + v_i\,, \nb\\
 h_{ij} &=& 2\psi \delta_{ij} + \left(\partial_i\partial_j-\frac{1}{3}\delta_{ij}\Delta\right)E \nonumber\\
 & & + \frac{1}{2}(\partial_iE_j+\partial_jE_i) + \gamma_{ij}\,,
\eqn
with $\Delta \equiv \delta^{ij}\partial_i \partial_j$ and  the constraints 
\bqn
\lb{eq3.3}
\partial^i v_{i} &=& \partial^i B_i = \partial^i E_i = 0\,,\nb\\
\partial^i \gamma_{ij} &=&  0\,, \quad \gamma^i_{\ i} = 0\,,
\eqn
where all the spatial indices are raised or lowered by $\delta^{ij}$ or $\delta_{ij}$, for example
$\partial^i v_{i} \equiv \delta^{ij}\partial_{j}v_{i}$,  and so on. Therefore, we have six scalars, $h_{00}$,  $w^0$, $B$, $v$, $\psi$ and $E$; three transverse vectors, $B_i$, $v_i$ and $E_i$; and one transverse-traceless tensor, $\gamma_{ij}$.  Under the following coordinate transformations, 
\bq
\lb{eq3.4}
t' = t + \xi^0\,, \quad {x'}^{i} = x^i +  \xi^{i} + \partial^i\xi\,,
\eq
where $ \partial_i\xi^{i} = 0$, these quantities  change as
\bqn
\lb{eq3.5a}
h_{00}'  &=& h_{00} - 2 \dot{\xi}^0\,, \quad \quad {w'}^0 = w^0 + \dot{\xi}^0\,, \nb \\
E' &=& E + 2\xi\,, \quad \psi' = \psi + \xi^0+\frac{1}{3}\Delta\xi\,, \quad v' = v + \dot{\xi}\,, \nb\\
B' &=& B - \xi^0 + \dot{\xi}\,, \\
\lb{eq3.5b}
B'_{i} &=& B_i + \dot{\xi}_i\,, \quad E'_{i} = E_i + 2{\xi}_i\,,\nb\\
v'_{i} &=& v_i + \dot{\xi}_i\,, \\
\lb{eq3.5c}
\gamma'_{ij} &=&   \gamma_{ij}\,.
\eqn

For the scalar part, let us choose the gauge 
\bq
\lb{eq3.6}
E = B = 0\,,
\eq
which are equivalently to choose the arbitrary functions $\xi^0$ and $\xi$ as $\xi = -E/2$ and $\xi^0 =B+\dot{\xi}$, so that the gauge freedom is completely fixed \footnote{In \cite{Foster06a}, the gauge $v = B = 0$ was adopted. However, as it can be seen from Eq.(\ref{eq3.5a}), in this case $\xi$ is fixed up to an arbitrary function $\hat{\xi}\left(x^k\right)$, that is, $\xi = \hat{\xi}\left(x^k\right) - \int{v dt}$, while $\xi^0$ is completely fixed by $\xi^0 = B+\dot{\xi}$.}.  Then, integrating out the variables $h_{00}$, $w^0$ and $v$,
we find that the quadratic action of the scalar part takes the form,
\bqn
\lb{eq3.7}
S_{\ae}^{(2, S)} &=& \frac{1}{8\pi G_{\ae}} \int{d^4x\left[\frac{\left(1-c_{13}\right)\left(2 + c_{13} + 3c_2\right)}{c_{123}}\dot{\psi}^2 \right.}\nb\\
&& ~~~~~~~~~~~~~~~~~~~~ \left. + \frac{2-c_{14}}{c_{14}}\psi\Delta \psi\right]\,.
\eqn
Thus, the ghost-free condition requires 
\bq
\lb{cd1}
q_S \equiv \frac{\left(1-c_{13}\right)\left(2+ c_{13} + 3c_2\right)}{c_{123}} > 0\,.
\eq
Then, the variation of $S_{\ae}^{(2, S)}$ with respect to $\psi$ yields the field equation, $\ddot{\psi} - c_S^2 \Delta \psi = 0$, where
\bqn
\lb{2.3af1}
c_S^2 \equiv \frac{c_{123}(2-c_{14})}{c_{14}(1-c_{13}) (2+c_{13} + 3c_2)}\,.
\eqn

For the vector part, we choose the gauge $\xi_i = -E_i/2$, so that $E_i ' = 0$. Then, after integrating out $B_i$, we find that the quadratic action of the vector part takes the form,
\bqn
\lb{eq3.8}
S_{\ae}^{(2, V)} &=& \frac{1}{16\pi G_{\ae}} \int{d^4x\Bigg[c_{14}\dot{v}^i\dot{v}_i }\nb\\
&& ~~~~~~~~~~~~ + \frac{2c_1 - c_{13}c_{-}}{2(1-c_{13})}v^i\Delta v_i\Bigg]\,.
\eqn
  Clearly, the ghost-free condition of the vector part now requires 
\bq
\lb{cd2}
 q_V \equiv c_{14} > 0\,.
\eq
Then, the variation of $S_{\ae}^{(2, V)}$ with respect to $v_i$ yields the field equation,
$\ddot{v}_i - c_V^2 \Delta v_i = 0$,
where
\bqn
\lb{2.3af2}
c_V^2 &\equiv& \frac{2c_1 -c_{13} c_{-}}{2c_{14}(1-c_{13})}\,.
\eqn

Similarly, the quadratic action of the tensor part takes the form,
\bqn
\lb{eq3.7}
S_{\ae}^{(2, T)} &=& \frac{1}{64\pi G_{\ae}} \int{d^4x\Big[\left(1-c_{13}\right)\dot{\gamma}^{ij}\dot{\gamma}_{ij} + \phi^{ij}\Delta \gamma_{ij}\Big]}\,.  \nonumber\\
& &
\eqn
Thus, the ghost-free condition of the tensor part requires 
\bq
\lb{cd3}
q_T \equiv  1-c_{13} > 0\,.
\eq
Then, the variation of $S_{\ae}^{(2, T)}$ with respect to $\gamma_{ij}$ yields the field equation, $\ddot{\gamma}_{ij} - c_T^2 \Delta \gamma_{ij} = 0$,
where
\bqn
\lb{2.3af3}
c_T^2 = \frac{1}{1-c_{13}}\,.
\eqn


\end{document}